\baselineskip=18pt
\overfullrule=0pt
\font\cs=cmr12 scaled\magstep2

\magnification 1200
\null
\footline={\ifnum\pageno>0 \hfil \folio\hfil \else\hfill \fi}
\pageno=0

\centerline{}

\vskip 20pt
\centerline{\bf {\cs A CONSISTENT COMPUTATIONAL}}
\vskip 20pt
\centerline{\bf {\cs TIME-DEPENDENT ELECTRON-EXCHANGE THEORY}}
\vskip 20pt
\centerline{\bf {\cs WITH NON-REDUNDANT TIME EVOLUTION}}

\vskip 20pt

\centerline{\bf Charles A. Weatherford}
\vskip 10pt
\centerline{Department of Physics}
\centerline{Florida A$\&$M University, Tallahassee, FL 32307}

\vskip 20pt

\centerline{\bf Abstract}

\vskip 15pt

\noindent
In the present work, a new time-dependent exchange theory is presented
wherein the symmetry constraints, on a multi-electron  wavefunction, are
properly accounted for.  In so doing, the equations of motion, incorporating 
the required symmetry, are derived and a solution algorithm employing an 
implicit split-operator procedure is described.  A technique 
(using an orthonormalization transformation and a unitary rotation), for
explicitly enforcing the required constraints, which render the computations
tractible and provide for non-redundant time evolution, is also presented.  This
amounts to the calculation of the appropriate numerically determined
guage.  The invariance of the derived orbital equations of motion with respect to the
transformations is explicitly demonstrated.

\vskip 20pt

\item\item{PACS:}34.10.+x; 31.25.-v; 31.15.NE; 31.70.Hq
\item\item{Keywords:}Electron-electron correlation; Two-electron
systems; Time-dependent exchange

\vskip 40pt

\par\vfill\eject

\centerline{\bf I.\ INTRODUCTION}

\vskip 20pt

The time-dependent Schr\"odinger equation (TDSE) 

$$
i  {{d}\over{dt}} \Psi = {\hat H} \Psi
\eqno(1)
$$

\noindent
describes the dynamics of quantum mechanical 
systems, and in particular
is applicable to systems consisting 
of $N$-electrons. The present application
is restricted to atomic and molecular systems with
all nuclei fixed in space 
(assumed infinitely massive). Note that atomic
units are used throughout this work.[1] In these 
units, $\hbar = m_e = e = 1$, where $\hbar$ is Planck's
constant divided by $2\pi$, $m_e$ is the electron mass,
and $e$ is the electron charge.  $\Psi$ is the system
wavefunction and is a function of time ($t$) and of the
coordinates of the particles making up the system.
${\hat H}$ is the Hamiltonian operator and consists
of the kinetic energy operators of all of the particles
in the system, plus the interaction potential between
the particles as well as any external potential. In
general, ${\hat H}$  might explicitly depend on time
through the external interaction potential, and while this
would present no fundamental complication, the present
work assumes ${\hat H}$ does not explicitly depend on
time. Also, in order to make the essential points of the
present work, the number of electrons will be restricted
to two. This would seem at first to be a drastic reduction
in complexity, which of course it is, but the essential
points can be made most clearly for two electron systems,
and indeed, two electron systems, such as the hydrogen molecule
($H_2$), electron-hydrogen atom ($e+H$) scattering, and the helium
atom ($He$), are important systems.

The TDSE represents an initial value 
problem such that if the
value of the wavefunction at $t=0$ is specified, and if the
TDSE can be accurately solved, then the quantum mechanical
dynamics will be encoded in the solution (wavefunction) at 
$t=\infty$ or an approximation thereunto.  In section II, 
the multiconfigurational time-dependent exchange theory
will be presented in a form which is specialized to two-electron
systems. In addition, since much of the present development is herein
presented for the first time, only one orbital per particle
will explicitly be considered. It should be noted that the present formulation
can be viewed as a modification of the multiconfigurational time-dependent
Hartree (MCTDH) theory of Manthe, 
Meyer, and Cederbaum,[2,3] appropriate for fermions, and as such, it
may be generalized to several fermions
and multi-function representations of each one. 
But in addition, the present paper presents an
explicit prescription for enforcing the required
constraints on the time-dependent orbitals such
that \lq\lq non-redundant time evolution\rq\rq\ is 
guaranteed--this appears to be a new contribution, along with
the new explicit treatment of the exchange symmetry and the very
compact form of the EOM. 
Section II.A derives the equations
of motion (EOM) assuming orbital orthonormality and
non-redundant time evolution; Section II.B presents the
derivation of the transformations--(1) an orthonormalization matrix
and (2) a unitary rotation matrix, which is the solution
of a first order differential equation in time.  The
application of these two orbital transformations 
results in orbital orthonormality and non-reduntant
orbital time evolution; Section II.C demonstrates the
invariance of the EOM with respect to the two
transformations; Section
III describes a solution algorithm using an implicit
split-operator procedure (ISOP)[4,5].  Finally, section IV will
present the conclusions.

\vskip 30pt

\centerline{\bf II.\ TIME-DEPENDENT ELECTRON EXCHANGE THEORY}

\vskip 20pt

\item{II.A} Equations of Motion

\vskip 10pt

As specified above, the present development is applied to two-electron
systems and thus the TDSE is represented as

$$
i  {{d}\over{dt}} \Psi(1,2) = {\hat H(1,2)} \Psi(1,2)
\eqno(2)
$$

\noindent
where the notation for $\Psi$ stands for $\Psi(1,2) =
\Psi({\vec r_1}, {\vec r_2}; t)$, and ${\hat H}(1,2) =
{\hat H}({\vec r_1}, {\vec r_2})$ such that the time dependence
is assumed for all wavefunctions and orbitals (to be defined
below)--the Hamiltonian however is assumed to have no explicit time
dependence.  	For a two-electron system, 
the wavefunction factors
into a space part times a spin part and 
the total wavefunction must be
completely antisymmetric with respect to 
electron exchange.[6] The spin
states are either singlets or triplets.  
The Hamiltonian is spin
independent at the level of theory under 
consideration. More explicitly,
the Hamiltonian is given by

$$
{\hat H}(1,2) = {\hat H}_0(1) + {\hat H}_0(2) + V(1,2) =
{\hat H}_0(1, 2) + V(1,2)
\eqno(3)
$$

\noindent
where ${\hat H}_0(j) = {\hat T}_0(j) + V_0(j)$ and
${\hat T}_0(j) = - {{1}\over{2}} \nabla^2_{{\vec r}_j}$.  Note that
if $V(1,2)$ is symmetric, then ${\hat H}(1,2)$ is also.

For a two-electron system, the spatial wavefunction, corresponding to
a spin singlet ($\Psi^{(+)}$), is symmetric with respect to exchange,
while the spatial part is antisymmetric for a spin triplet
($\Psi^{(-)}$).  The spatial wavefunction then depends on the spatial parts
of the two electrons and on time.  It resides in a six dimensional ($6D$)
coordinate space.  If it were practical, the full six spatial
dimensional wavefunction would be propagated in time. However, this
is not practical at the present stage of computer technology--it
may be in the near future however.  Certainly, for systems composed
of three or more electrons, the full dimensional solution is
not available and will not be for the forseeable future.  Thus, a
decomposition into a direct product of three 
dimensional ($3D$) subspaces (one
for each electron) is the advisable procedure.  
Then, the spatial part of the two-electron
wavefunction may then be expanded, at the minimal expansion length, 
in the manner of [2,3], as

$$
\eqalign{
\Psi^{(+)}(1,2) =& \sum_{j=1}^3  A_j^{({\phi},{+})} \Phi_j^{({\phi},{+})}(1,2) \cr
\Psi^{(-)}(1,2) =& A_1^{({\phi}, {-})} \Phi_1^{({\phi}, {-})}(1,2) \cr}
\eqno(4)
$$

\noindent
The $A$'s are purely time-dependent coefficients multiplying each
two-electron configuration function. The superscript ${({\phi},{\pm})}$
indicates that the one-electron orbitals, labeled by $\phi$, are 
used, and the permutation
symmetry is singlet ($+$) or triplet ($-$).
Thus, the singlet 
spatial wavefunction is a superposition of three two-electron
configurations $\Phi_j^{(\phi, +)}$ and the triplet has one two-electron configuration
function $\Phi_1^{(\phi, -)}$.  The minimal length of these expansions is determined
by the required invariance of $\Psi^{(\pm)}$ with respect to 
arbitrary rotations among the
one-electron orbitals[2,3] comprising the 
$\Phi_1^{(\phi, \pm)}$'s (see below).  
$\Phi_j^{(\phi, +)}: j=1,2,3$ is
symmetric with respect to electron exchange 
and $\Phi^{(\phi, -)}$ is antisymmetric with respect
to electron exchange.  There are two one-electron orbitals required
to describe the two-electron configurations: they are
labeled $\phi_j^{(\pm)}: j=1,2$.

$$
\eqalign{
\Phi_1^{(\phi, +)}(1,2) =& \phi_1^{(+)}(1) \phi_1^{(+)}(2) \cr
\Phi_2^{(\phi, +)}(1,2) =& {{1}\over{\sqrt{2}}} 
       \big[ \phi_1^{(+)}(1) \phi_2^{(+)}(2) 
      + \phi_1^{(+)}(2) \phi_2^{(+)}(1) \big] \cr
\Phi_3^{(\phi, +)}(1,2) =& \phi_2^{(+)}(1) \phi_2^{(+)}(2) \cr
\Phi_1^{(\phi, -)}(1,2) =& {{1}\over{\sqrt{2}}} 
       \big[ \phi_1^{(-)}(1) \phi_2^{(-)}(2) 
      - \phi_1^{(-)}(2) \phi_2^{(-)}(1) \big] \cr}
\eqno(5)
$$

Consistent with the decomposition of the $6D$ space into a 
direct product of $3D$ subspaces, the objective is to derive
two coupled time-dependent $3D$ equations for the one-electron
orbitals.  In addition, time-dependent equations for the purely time-dependent
coefficients need to be derived (three for the singlet and one
for the triplet--see Eq. (4)). 

Now in general, the $\phi$'s are not 
necessarily computationally orthonormal
because of numerical inaccuracies.  If Eqs. (5) are substituted into Eqs. (4),
which are then substituted into Eq. (2), aside from nonzero off-diagonal,
and non-unit diagonal, overlaps
$S_{ij}^{(\phi, \pm)} =\  <\phi_{i}^{(\pm)} \vert \phi_{j}^{(\pm)} >$, a set of
nonzero \lq\lq derivative overlaps\rq\rq\ given by 
$D_{ij}^{(\phi, \pm)} = <\phi_{i}^{(\pm)} \vert {\dot \phi}_{j}^{(\pm)} >$,
where the \lq over-dot\rq\ represents a time derivative,
will appear in the equations.  One of the principal distinctions
of the MCTDH theory of [2,3] is the use of the purely time-dependent
coefficients, as indicated in Eqs. (4), as contrasted with, for example,
a time-dependent Hartree-Fock theory (TDHF) [7], wherein
the time-dependence is described solely by the one-electron
orbitals.  As pointed out
in [2,3], this additional set of time-dependent 
coeficients produces a redundant
description which allows for the incorporation 
of certain constraints.  To see this,
consider a new set of one-electron orbitals, 
$\psi_j^{(\pm)}$, related to the old set by

$$
\eqalign{
\psi_1^{(\pm)} =& \phi_1^{(\pm)} b_{11}^{(\pm)}  + \phi_2^{(\pm)} b_{21}^{(\pm)}  \cr
\psi_2^{(\pm)} =& \phi_1^{(\pm)} b_{12}^{(\pm)}  + \phi_2^{(\pm)} b_{22}^{(\pm)}  \cr}
\eqno(6)
$$  

\noindent
Assuming ${\bf b}$ is unitary and purely time-dependent,
(${\bf b}{\bf b}^{\dagger} = {\bf b}^{\dagger} {\bf b} = {\bf 1}$), 
this can be written

$$
\eqalign{
{\tilde {\vec \psi}}^{(\pm)} =&  {\tilde {\vec \phi}}^{(\pm)} {\bf b}^{(\pm)} \cr
 {\vec \psi}^{(\pm)} =&  {\tilde {\bf b}}^{(\pm)} {\vec \phi}^{(\pm)}  \cr}
\eqno(7)
$$

\noindent
where ${\dagger}$ indicates hermitian conjugate. 
Assuming Eq. (7) can be inverted (where the tilde indicates vector
or matrix
transpose) as per

$$
\eqalign{
{\tilde {\vec \phi}}^{(\pm)} =&  {\tilde {\vec \psi}}^{(\pm)} {\bf {b}^{\dagger}}^{(\pm)} \cr
 {\vec \phi}^{(\pm)} =&   {\bf {b}^{*}}^{(\pm)} {\vec \psi}^{(\pm)}  \cr}
\eqno(8)
$$

\noindent
and inserted into Eqs. (5) and then into Eqs. (4), it will be see 
that Eqs. (4) can be written as

$$
\eqalign{
\Psi^{(+)}(1,2) =& 
  \sum_{j=1}^3 A_j^{(\psi,+)} \Phi_j^{(\psi, +)}(1,2) \cr
\Psi^{(-)}(1,2) =& 
  A_1^{(\psi, -)} \Phi_1^{(\psi, -)}(1,2) \cr}
\eqno(9)
$$

\noindent
where

$$
\eqalign{
\Phi_1^{(\psi, +)}(1,2) =& \psi_1^{(+)}(1) \psi_1^{(+)}(2) \cr
\Phi_2^{(\psi, +)}(1,2) =& {{1}\over{\sqrt{2}}} 
       \big[ \psi_1^{(+)}(1) \psi_2^{(+)}(2) 
      + \psi_1^{(+)}(2) \psi_2^{(+)}(1) \big] \cr
\Phi_3^{(\psi, +)}(1,2) =& \psi_2^{(+)}(1) \psi_2^{(+)}(2) \cr
\Phi_1^{(\psi, -)}(1,2) =& {{1}\over{\sqrt{2}}} 
       \big[ \psi_1^{(-)}(1) \psi_2^{(-)}(2) 
      - \psi_1^{(-)}(2) \psi_2^{(-)}(1) \big] \cr}
\eqno(10)
$$

\noindent
and

$$
\eqalign{
A_1^{(\psi, +)} =& A_1^{(\phi, +)} {{b_{11}^{(+)}}^*} {{b_{11}^{(+)}}^*}+ 
    A_2^{(\phi, +)} \sqrt{2}\ {b_{11}^{(+)}}^* {b_{21}^{(+)}}^* +
    A_3^{(\phi, +)} {b_{21}^{(+)}}^* {b_{21}^{(+)}}^*\cr
A_2^{(\psi, +)} =&  A_1^{(\phi, +)} \sqrt{2}\ {b_{11}^{(+)}}^* {b_{12}^{(+)}}^* + 
    A_2^{(\phi, +)} \big[ {b_{11}^{(+)}}^* {b_{22}^{(+)}}^* + 
            {b_{21}^{(+)}}^* {b_{12}^{(+)}}^* \big] +
    A_3^{(\phi, +)} {b_{22}^{(+)}}^* {b_{21}^{(+)}}^*\cr
A_3^{(\psi, +)} =&  A_1^{(\phi, +)} {b_{12}^{(+)}}^* {b_{12}^{(+)}}^* + 
    A_2^{(\phi, +)} \sqrt{2}\ {b_{12}^{(+)}}^* {b_{22}^{(+)}}^* +
    A_3^{(\phi, +)} {b_{22}^{(+)}}^* {b_{22}^{(+)}}^* \cr
A_1^{(\psi, -)} =&  
    A_1^{(\phi, -)} \big[ {b_{11}^{(-)}}^* {b_{22}^{(-)}}^* - 
            {b_{21}^{(-)}}^* {b_{12}^{(-)}}^* \big]
    \cr }
\eqno(11)
$$

\noindent
The significant point is that the four purely time-dependent
matrix elements defining ${\bf b}^{(\pm)}$ are completely
arbitrary because of the invariance of Eqs. (9) and (10) with
respect to the transformation described by Eq. (7).

The utility of the arbitrariness of ${\bf b}^{(\pm)}$  lies in
its use to fix four matrix elements involving the $\psi$'s.
As a rationale for this, consider that there are four overlaps  
$\ S_{ij}^{(\psi, \pm)} =\  <\psi_{i}^{(\pm)} \vert \psi_{j}^{(\pm)} >$
and four derivative overlaps
$D_{ij}^{(\psi, \pm)} = \ <\psi_{i}^{(\pm)} \vert {\dot \psi}_{j}^{(\pm)} >$.
Thus there are eight such matrix elements.  However, only
four are independent.  If the four choices are made corresponding
to $D_{11}^{(\psi, \pm)}=0, D_{21}^{(\psi, \pm)}=0, 
D_{22}^{(\psi, \pm)}=0$, and $S_{21}^{(\psi, \pm)}=0$, then it can be
easily shown that the other four matrix elements are fixed.  Thus,
$D_{i,j}^{(\psi, \pm)} = 0, 
S_{i,j}^{(\psi, \pm)} = \delta_{i,j}$ for all $i,j$.  Actually, the
diagonal overlaps are arbitrary constants which may be set to one.
In section II.B below, a systematic way of implementing these
constraints will be given.  In so doing, it will be seen that
the enforcement of orthonormality and non-reduntant time-evolution,
must be done in two different transformation 
steps, separated in sequence by
the time propagation over the time interval $\Delta t$.

The derivation of the EOM proceeds by a 
projection method.  A critical property in the derivation by
projection of the EOM for the $A$'s is the orthonormality
of the two-electron configuration functions such that

$$
<< \Phi_i^{(\psi, \pm)} \vert \Phi_j^{(\psi, \pm)} >> = \delta_{ij},
\eqno(12)
$$

\noindent
where the double brackets $<< \vert >>$ represents 6D integration
over the coordinates of both particles, and the zero values
of the derivative overlap matrix elements of the two-electron
configuration functions

$$
<< \Phi_i^{(\psi, \pm)} \vert {\dot \Phi}_j^{(\psi, \pm)} >> = 0.
\eqno(13)
$$

\noindent
These relations follow immediately
from the orthonormality of the $\psi$'s and the zero values
of the one-electron derivative 
overlap matrix elements.  The procedure is thus to
substitute Eqs. (10) into Eqs. (9), and then to use the result
in Eq. (2).  Then project from the left by ${\Phi_k^{(\psi, \pm)}}^*$
and integrate over the coordinates of particles one and two.  The result
is

$$
{\dot A}_k^{(\psi, \pm)} = - i \sum_{j=1}^{N^{(\pm)}} 
     << {\Phi_k^{(\psi, \pm)}} \vert {\hat H}
        \vert {\Phi_j^{(\psi, \pm)}} >> {A_j^{(\psi, \pm)}}
\eqno(14)
$$

\noindent
where ${N^{(\pm)}} = 3/1$.  for the singlet case ($+$ sign), $j,k:1,2,3$.
For the triplet case, ($-$ sign), $j,k:1$.  

In order to derive the
EOM for the time-dependent one-electron orbitals, the singlet
and triplet cases are considered separately.
To derive EOM for the one-electron orbitals, a similar projection
(as above for the $\Phi$'s)
is employed, except now the one-electron orbitals are used.
First, project from the left by ${\psi_1^{(+)}}(1)^*$ on Eq. (2)
and integrate over the coordinates of particle one.  Then use the
overlap and derivative overlap constraints (in the $\psi$-basis) to obtain

$$
\eqalign{
A_1^{(\psi, +)} {\dot \psi}_1^{(+)}(2) 
      +& {{1}\over{\sqrt{2}}} A_2^{(\psi, +)} {\dot \psi}_2^{(+)}(2) 
      + {\dot A}_1^{(\psi, +)} {\psi}_1^{(+)}(2) 
      + {{1}\over{\sqrt{2}}} {\dot A}_2^{(\psi, +)} {\psi}_2^{(+)}(2) \cr
=& - i \sum_{j=1}^3 
      < \psi_1^{(+)}(1) \vert {\hat H} \vert \Phi_j^{(\psi, +)}(1,2) >_1
      A_j^{(\psi, +)} \cr}
\eqno(15)
$$

\noindent
where $<\vert \vert>_1$ indicates an integral over the coordinates
of particle one.

Secondly, project from the left by ${\psi_2^{(+)}}(2)^*$ on Eq. (2)
and integrate over the coordinates of particle two.  Then use the
overlap and derivative overlap constraints, again, to obtain

$$
\eqalign{
{{1}\over{\sqrt{2}}} A_2^{(\psi, +)} {\dot \psi}_1^{(+)}(1) 
      +& A_3^{(\psi, +)} {\dot \psi}_2^{(+)}(1) 
      + {{1}\over{\sqrt{2}}} {\dot A}_2^{(\psi, +)} {\psi}_1^{(+)}(1) 
      + {\dot A}_3^{(\psi, +)} {\psi}_2^{(+)}(1) \cr
=& - i \sum_{j=1}^3 
      < \psi_2^{(+)}(2) \vert {\hat H} \vert \Phi_j^{(\psi, +)}(1,2) >_2
      A_j^{(\psi, +)} \cr}
\eqno(16)
$$

\noindent
where $<\vert \vert>_2$ indicates an integral over the coordinates
of particle two.

Exactly the same two projections are done for the triplet
case as was done for the singlet case.  The results are (in analogy
to Eq. (15))

$$
{{1}\over{\sqrt{2}}} A_1^{(\psi, -)} {\dot \psi}_2^{(-)}(2) 
      + {{1}\over{\sqrt{2}}} {\dot A}_1^{(\psi, -)} {\psi}_2^{(-)}(2)
= - i < \psi_1^{(-)}(1) \vert {\hat H} \vert \Phi_1^{(\psi, -)}(1,2) >_1
      A_1^{(\psi, -)}
\eqno(17)
$$

\noindent
and (in analogy to Eq. (16))

$$
{{1}\over{\sqrt{2}}} A_1^{(\psi, -)} {\dot \psi}_1^{(-)}(1) 
      + {{1}\over{\sqrt{2}}} {\dot A}_1^{(\psi, -)} {\psi}_1^{(-)}(1)
= - i < \psi_2^{(-)}(2) \vert {\hat H} \vert \Phi_1^{(\psi, -)}(1,2) >_2
      A_1^{(\psi, -)}
\eqno(18)
$$

\noindent
The three EOM are thus given by Eqs. (14),(15), and (16) for the
singlet case, and by Eqs. (14),(17), and (18) for the triplet case.

Plugging Eqs. (10) into Eqs. (15) and (16), then plugging
Eqs. (10) into Eqs. (17) and (18), and then reversing the 
coordinate labels in Eq. (15) and Eq. (17) 
(e.g. $1 \leftrightarrow 2$), and 
finally writing in matrix form, results in

$$
{\dot {\vec \Psi}}^{(\psi, \pm)}(1)
   = - i\ {\hat {\bf h}}^{(\psi, \pm)}(1)\ {\vec \Psi}^{(\psi, \pm)}(1)
\eqno(19)
$$

\noindent
where the symbols are defined differently for the singlet
and triplet cases, and where

$$
{\vec \Psi}^{(\psi, \pm)}(1)
   = {\bf a}^{(\psi, \pm)}\ {\vec \psi}^{(\pm)}(1)
\eqno(20)
$$

\noindent
Now, the $h$ and $a$-matrices are defined as

$$
{\hat {\bf h}}^{(\psi, \pm)}(1) =
\left(\matrix{
{\hat h}_{11}^{(\psi, \pm)}(1) &  {\hat h}_{12}^{(\psi, \pm)}(1) \cr
\noalign{\bigskip}
 {\hat h}_{21}^{(\psi, \pm)}(1) & {\hat h}_{22}^{(\psi, \pm)}(1) \cr} \right),
\eqno(21)
$$

\noindent
and, for the singlet

$$
{\bf a}^{(\psi, +)} = \left(
\matrix
    {A_1^{(\psi, +)} & {{1}\over{\sqrt{2}}} A_2^{(\psi, +)}  \cr
    {{1}\over{\sqrt{2}}} A_2^{(\psi, +)} & A_3^{(\psi, +)}  \cr}
                \right),
\eqno(22)
$$

\noindent
and for the triplet

$$
{\bf a}^{(\psi, -)} = \left(
\matrix{
     0 &  A_1^{(\psi, -)}  \cr
    - A_1^{(\psi, -)} & 0  \cr}
                \right).
\eqno(23)
$$

\noindent
Finally, the $h_{k,j}^{(\psi, \pm)}$ matrix elements are defined by

$$
{\hat h}_{kj}^{(\psi, \pm)}(1) = < \psi_k^{(\pm)}(2) \vert
    {\hat H}(1,2) \vert \psi_j^{(\pm)}(2) >_2
\eqno(24)
$$

\noindent
From here on, the actual orbital coordinate labels
will be dropped, except in several circumstances
where it is convenient to exhibit them for clarity.

\vskip 20pt

\item{II.B} Orbital Transformations

\vskip 10pt

In the derivation of the EOM [Eqs. (14,19)] using the $\psi$-basis,
it has been assumed that ${\bf S}^{(\psi, \pm)} = {\bf 1}$ and 
 ${\bf D}^{(\psi, \pm)} = {\bf 0}$.  However, just because these
two assumptions have been made in the deriving the EOM, does
not automatically result in the enforcement of the constraints
expressed in the two assumptions.  It is well known that if
the multielectron time-dependent Schr\"odinger equation
is solved exactly as an $N$-electron problem, without
the orbital direct product ansatz, and if the wave function
at $t=0$ is orthonormal, then, in principle, the wave function
should remain orthonormal for all time.  However, computational
errors will inevitably accrue and destroy this orthonormality.
For an orbital direct product decomposition, this
orthonormality assumption is in principal, still valid,
subject to numerical inaccuracies. If this orbital
orthonormality is lost because of numerical inaccuracies,
it is not sufficient to just re-orthonormalize without
appropriately modifying the other terms which appear
in the EOM--in fact, the EOM should be invariant with
respect to this orthonormalization transformation.
It is one of the salient features of the present work,
that this invariance is explicit for the EOM derived herein.
This invariance is demonstrated below.  Also, a
procedure for enforcement is described.

It can be easily shown that if in some basis ($\psi$ for example)
${\bf S^{(\psi, \pm)}} = {\bf 1}$, then the $D$-matrix is
anti-hermitian (e.g. ${\bf D}^{(\psi, \pm)} + 
{\bf D}^{{(\psi, \pm)}^{\dagger}} = {\bf 0}$).  On the other
hand, if ${\bf D}^{(\psi, \pm)} = {\bf 0}$, then 
${\bf D}^{{(\psi, \pm)}^{\dagger}} = {\bf 0}$ and 
${\bf S^{(\psi, \pm)}} = {\bf C}^{(\psi, \pm)}$ where 
${\bf C}^{(\psi, \pm)}$
is a constant matrix, not necessarily ${\bf 1}$.  As indicated
above, however, orthonormality and a null $D$-matrix are
consistent with each other.

\vskip 5pt

\item\item{II.B-1} Orthonormalization Transformation

\vskip 5pt

This section is concerned with the following overlap
matrices:

$$
\eqalign{
{\bf S}^{(\chi, \pm)} = 
    < {\tilde {\vec \chi}^{(\pm)}} \vert\  {\tilde {\vec \chi}^{(\pm)}} >  \cr
{\bf S}^{(\phi, \pm)} = 
    < {\tilde {\vec \phi}^{(\pm)}} \vert\  {\tilde {\vec \phi}^{(\pm)}} >  \cr }
\eqno(25)
$$

\noindent
A symmetric orthonormalization procedure [8] is utilized,
at the R-end of the $\Delta t$ interval
(note that the transformation matrix ${\bf X}^{(\pm)}$ is not unitary,
however, note that ${\bf X}^{(\pm)} = {\bf X}^{{(\pm)}^{\dagger}}$),
to go from the $\chi$-set to the $\phi$-set:

$$
\phi_{\mu}^{(\pm)} = \sum_{\nu} \ \chi_{\nu}^{(\pm)} X_{\nu , \mu}^{(\pm)}
\eqno(26)
$$

\noindent
or in matrix notation

$$\eqalign{
{\tilde {\vec \phi}^{(\pm)}} &= {\tilde {\vec \chi}^{(\pm)}}\  {\bf X}^{(\pm)} \cr
{\vec \phi}^{(\pm)} &= {\tilde {\bf X}^{(\pm)}}\ {\vec \chi}^{(\pm)} \cr}
\eqno(27)
$$

\noindent
where

$$
{\bf X}^{(\pm)} = {\bf U}^{(\pm)}\ 
\big[ {\bf {\vec s}}^{(\chi, \pm)} \big]^{-{{1}\over{2}}}\ {{\bf U}^{(\pm)}}^{\dagger}
\eqno(28)
$$

\noindent
and where ${\bf U}^{(\pm)}$ is the unitary matrix that diagonalizes
$\big[ {\bf S}^{(\chi, \pm)} \big]^{\pm 1}$ (the overlap matrix,
or its inverse, in the $\chi$-basis)

$$
{{\bf U}^{(\pm)}}^{\dagger}\ \big[ {\bf S}^{(\chi, \pm)}\big]^{\pm 1}\ {\bf U}^{(\pm)} = 
\big[ {\bf {\vec s}}^{(\chi, \pm)} \big]^{\pm 1}
\eqno(29)
$$

\noindent
such that ${\bf {\vec s}}^{(\chi, \pm)}$ is the diagonal matrix
of eigenvalues and $\big[ {\bf {\vec s}}^{(\chi, \pm)} \big]^{-{{1}\over{2}}}$
is the diagonal matrix of one over the square root of the
eigenvalues. Note that in Eq. (29), the ${\pm 1}$ is independent
of the ${\pm}$ which appears in the superscript ${(\chi, \pm)}$. Thus,

$$
\eqalignno{
{\bf S}^{(\phi, \pm)} &= {\bf  1} =
{{\bf X}^{(\pm)}}^{\dagger}\ {\bf S}^{(\chi, \pm)}\ {\bf X}^{(\pm)}; &(30a) \cr
{{\bf X}^{(\pm)}}^{\dagger} {{\bf X}^{(\pm)}} &= 
       {{\bf X}^{(\pm)}} {{\bf X}^{(\pm)}}^{\dagger} = 
       {{\bf X}^{(\pm)}} {{\bf X}^{(\pm)}} = 
       \big[ {\bf S}^{(\chi, \pm)} \big]^{-1}. &(30b) \cr}
$$

\noindent
It can easily be seen that

$$
\eqalignno{
{{\bf X}^{(\pm)}}^{-1} &= {\bf U}^{(\pm)}\ 
\big[ {\bf {\vec s}}^{(\chi, \pm)} \big]^{{{1}\over{2}}}\ {{{\bf U}^{(\pm)}}^{\dagger}}; &(31a) \cr
{\tilde {\bf X}}^{{(\pm)}^{-1}} &= {{\bf U}}^{{(\pm)}^{*}}\ 
\big[ {\bf {\vec s}}^{(\chi, \pm)} \big]^{{{1}\over{2}}}\ {\tilde {{\bf U}}^{(\pm)}}. &(31b) \cr}
$$

\vskip 5pt

\item\item{II.B-2} Unitary Rotation Transformation

\vskip 5pt

In this section, it is assumed that the EOM in the ${\phi}$-set
has been solved over the interval $\Delta t$ so that the
$D$-matrix in the $\phi$-set can be calculated at
the R-end of the time interval.  Thus, this section is concerned 
with the following derivative overlap matrices calculated
at the R-end of the time interval:

$$
\eqalignno{
{\bf D}^{(\phi, \pm)} &= 
    < {\tilde {\vec \phi}}^{(\pm)} \vert\ {\tilde {\dot {\vec \phi}}^{(\pm)}} > ; &(32a) \cr
{\bf D}^{(\psi, \pm)} &= 
    < {\tilde {\vec \psi}^{(\pm)}} \vert\ {\tilde {\dot {\vec \psi}}^{(\pm)}} > . &(32b) \cr}
$$

\noindent
The objective is to find a unitary transformation (at the
R-end of the time interval) of the $\phi$-set to
the $\psi$-set, such that ${\bf D}^{(\psi, \pm)} = {\bf 1}$. Thus the
transformation is represented by

$$
\psi_{\mu}^{(\pm)} = 
  \sum_{\nu} \ \phi_{\nu}^{(\pm)} b_{\nu , \mu}^{(\pm)}
\eqno(33)
$$

\noindent
or in matrix notation, this is described by Eq. (7) above.
Note that the overlaps, given by this transformation, are
related by

$$
{\bf S}^{(\psi, \pm)} =
{{\bf b}^{(\pm)}}^{\dagger}\ {\bf S}^{(\phi, \pm)}\ {\bf b}^{(\pm)}
\eqno(34)
$$

\noindent
Clearly, if the $\phi$-set is orthonormal, and if ${\bf b}$ is
unitary, then the $\psi$-set is orthonormal.

Now, if Eq. (33) is used in Eq. (32), then

$$
{\bf D}^{(\psi, \pm)} = {\bf b}^{{(\pm)}^{\dagger}}
\bigg[
{\bf D}^{(\phi, \pm)} {\bf b}^{(\pm)} + {\bf S}^{(\psi, \pm)} {\dot {\bf b}}^{(\pm)}
\bigg]
\eqno(35)
$$

\noindent
If we demand ${\bf D}^{(\psi, \pm)} = {\bf 0}$, and assuming
orthonormality of the $\psi$-set, then

$$
{\dot {\bf b}}^{(\pm)} +
{\bf D}^{(\phi, \pm)} {\bf b}^{(\pm)}
 = {\bf 0}
\eqno(36)
$$

\noindent
Now ${\bf D}^{(\phi, \pm)}$ is anti-hermitian since the $\psi$-set
is orthonormal, as can be seen from Eq. (34), given the
orthonormality of the $\phi$-set. It is then known [7] that use of the Cayley
decomposition to propagate ${\bf b}^{\pm}$ in time, will
preserve unitarity.  Therefore, if Eq. (36) is solved, using ${\bf D}^{(\phi, \pm)}$
at the R-end, then ${\bf b}^{(\pm)}$ will be calculated at the R-end
such that ${\bf D}^{(\psi, \pm)} = {\bf 0}$ at the R-end.

\vskip 20pt

\item{II.C} EOM Invariance

\vskip 10pt

The objective of this section is to show the
invariance of the EOM with respect to the
orthonormalization transformation ($\bf X$) and the
rotation matrix ($\bf b$).  This will be done by
first surmising a compact form for the full
two-electron wave function (see Eq. (2)), and then
deriving the EOM without assuming orthonormality or
null $D$-matrices, for each of the three bases ($\chi, \phi, \psi$).
It will then be shown that the resultant EOM are invariant
in form with respect to the linear transformations
given by ${\bf X}$ and ${\bf b}$.  Note that, from hereon, unless
otherwise noted,
the $\pm$ notation will be dropped with the understanding that
the distinction still applies.
Collecting some pertinent formulas for reference, we show
the following, which are inferred from Eqs. (4, 20, and 24) above:

$$
\eqalignno{
& \underline{\bf {\chi-set}} \cr
 \Psi_{\chi}(1,2) &= {\tilde {\vec \chi}}(1)\  
    {\bf a}^{(\chi)}\ {\vec \chi}(2)\  ; &(37a) \cr
{\vec \Psi}^{(\chi)} &= {\bf a}^{(\chi)} {\vec \chi}; &(37b) \cr
{\hat {\bf h}}^{(\chi)} &= < {\tilde {\vec \chi}} \vert {\hat H} \vert\ 
      {\tilde {\vec \chi}} > . &(37c) \cr 
& \underline{\bf {\phi-set}} \cr
\Psi_{\phi}(1,2) &= {\tilde {\vec \phi}}(1)\  
    {\bf a}^{(\phi)}\ {\vec \phi}(2)\  ; &(38a) \cr
{\vec \Psi}^{(\phi)} &= {\bf a}^{(\phi)} {\vec \phi}; &(38b) \cr
{\hat {\bf h}}^{(\phi)} &= < {\tilde {\vec \phi}} \vert {\hat H} \vert\ 
      {\tilde {\vec \phi}} > . &(38c) \cr 
& \underline{\bf {\psi-set}} \cr
\Psi_{\psi}(1,2) &= {\tilde {\vec \psi}}(1)\  
    {\bf a}^{(\psi)}\ {\vec \psi}(2)\  ; &(39a) \cr
{\vec \Psi}^{(\psi)} &= {\bf a}^{(\psi)} {\vec \psi}; &(39b) \cr
{\hat {\bf h}}^{(\psi)} &= < {\tilde {\vec \psi}} \vert {\hat H} \vert\ 
      {\tilde {\vec \psi}} > . &(39c) \cr }
$$

\noindent
The actual orbital transformations have been given by Eqs. (7, 27).  It
must be shown that the EOM are invariant with respect to the two
transformations.  To this end, the EOM are derived below without
assuming orthonormality or a null $D$-matrix.

Begin by substituting Eq. (37a) into Eq. (2), multiplying from the left
by ${\tilde {\vec \chi}}(1)$ and integrating over the coordinates
of ${\vec r}(1)$--the result is

$$
{\bf D}^{(\chi)} {\bf a}^{(\chi)} {\vec \chi} +
{\bf S}^{(\chi)} {\bf {\dot a}}^{(\chi)} {\vec \chi} +
{\bf S}^{(\chi)} {\bf a}^{(\chi)} {\dot {\vec \chi}} =
{\hat {\bf h}}^{(\chi)} {\bf a}^{(\chi)} {\vec \chi}
\eqno(40)
$$

\noindent
Now, define a transformation of orbitals

$$
\eqalign{
{\tilde {\vec \psi}} &\equiv {\tilde {\vec \chi}}\  {\bf C} \cr
{\vec \psi} &= {\tilde {\bf C}}\  {\vec \chi} \cr}
\eqno(41)
$$

\noindent
where

$$
\eqalignno{
{\bf C}^{(\pm)} 
     &\equiv {\bf X}^{(\pm)} {\bf b}^{(\pm)} ; &(42a) \cr
{\bf C}^{(\pm)^*} 
     &= {\bf X}^{(\pm)^*} {\bf b}^{{(\pm)}^*} ; &(42b) \cr
{\bf C}^{(\pm)^{\dagger}} 
     &= {\bf b}^{{(\pm)}^{\dagger}}  {\bf X}^{(\pm)^{\dagger}} ; &(42c) \cr
{\bf C}^{(\pm)^{-1}} 
     &= {\bf b}^{{(\pm)}^{\dagger}}  {\bf X}^{(\pm)^{-1}} ; &(42d) \cr
{\tilde {\bf C}}^{(\pm)^{-1}} 
     &= {\tilde {\bf X}}^{(\pm)^{-1}} {\bf b}^{{(\pm)}^*} . &(42e) \cr }
$$

\noindent
Thus the transformation defined by ${\bf C}$ combines the orthonormalization
transformation with the rotation transformation.  In this manner,
the $\phi$-set of orbitals may be dispensed with.
Inverting Eqs. (41) results in

$$
\eqalign{
{\tilde {\vec \chi}} &\equiv {\tilde {\vec \psi}}\  {\bf C}^{-1} \cr
{\vec \chi} &= {\tilde {\bf C}}^{-1}\  {\vec \psi} \cr}
\eqno(43)
$$

\noindent
Then taking the time derivative of Eqs. (43), gives

$$
\eqalign{
{\dot {\tilde {\vec \chi}}} &\equiv {\tilde {\dot {\vec \psi}}}\  {\bf C}^{-1} +
        {\tilde {\vec \psi}}\  {\dot {\bf C}}^{-1}\cr
{\dot {\vec \chi}} &= {\dot {\tilde {\bf C}}}^{-1}\  {\vec \psi} + 
        {\tilde {\bf C}}^{-1}\  {\dot {\vec \psi}} \cr}
\eqno(44)
$$

\noindent
Substituting Eqs. (41), (43), and (44) into Eq. (40) gives

$$
\eqalign{
{\bf C}^{{\dagger}^{-1}} & 
     < {\tilde {\vec \psi}} \vert
        \big[ {\tilde {\dot {\vec \psi}}}\  {\bf C}^{-1} +
        {\tilde {\vec \psi}}\  {\dot {\bf C}}^{-1} \big] >
            {\bf a}^{(\chi)}\ {\tilde {\bf C}}^{-1} \ {\vec \psi} + 
{\bf C}^{{\dagger}^{-1}}  < {\tilde {\vec \psi}} \vert {\tilde {\vec \psi}} > 
        {\bf C}^{-1} {\dot {\bf a}}^{(\chi)}\ {\tilde {\bf C}}^{-1} \ {\vec \psi} +  \cr
{\bf C}^{{\dagger}^{-1}} & < {\tilde {\vec \psi}} \vert {\tilde {\vec \psi}} > 
      {\bf C}^{-1} {\bf a}^{(\chi)}
          \big[ {\dot {\tilde {\bf C}}}^{-1}\  {\vec \psi} + 
        {\tilde {\bf C}}^{-1}  {\dot {\vec \psi}}\ \big] 
= -i\  {\bf C}^{{\dagger}^{-1}} 
      < {\tilde {\vec \psi}} \vert {\hat H} \vert {\tilde {\vec \psi}} >
         {\bf C}^{-1} {\bf a}^{(\chi)}\ {\tilde {\bf C}}^{-1} \ {\vec \psi}. \cr}
\eqno(45)
$$

\noindent
Now define

$$
{\bf a}^{(\psi)} \equiv {\bf C}^{-1} {\bf a}^{(\chi)} {\tilde {\bf C}}^{-1}.
\eqno(46)
$$

\noindent
and then

$$
{\dot {\bf a}}^{(\psi)} = 
   {\dot {\bf C}}^{-1} {\bf a}^{(\chi)} {\tilde {\bf C}}^{-1} +
   {\bf C}^{-1} {\dot {\bf a}}^{(\chi)} {\tilde {\bf C}}^{-1} +
   {\bf C}^{-1} {\bf a}^{(\chi)} {\dot {\tilde {\bf C}}}^{-1}.
\eqno(47)
$$

\noindent
Also, it is clear that

$$
{\hat {\bf h}}^{(\psi)}  =  {\bf C}^{\dagger}\ {\hat {\bf h}}^{(\chi)}\ 
   {\bf C}
\eqno(48)
$$

\noindent
Then multiply Eq. (45) from the left by ${\bf C}^{\dagger}$ and use
Eqs. (46) and (47) to produce 

$$
{\bf D}^{(\psi)} {\bf a}^{(\psi)} {\vec \psi} +
{\bf S}^{(\psi)} {\bf {\dot a}}^{(\psi)} {\vec \psi} +
{\bf S}^{(\psi)} {\bf a}^{(\psi)} {\dot {\vec \psi}} =
{\hat {\bf h}}^{(\psi)} {\bf a}^{(\psi)} {\vec \psi}
\eqno(49)
$$

\noindent
Note that 

$$
\eqalignno{
{\bf S}^{(\psi)} &= < {\tilde {\vec \psi}} \vert {\tilde {\vec \psi}} > ; &(50a) \cr
{\bf D}^{(\psi)} &= < {\tilde {\vec \psi}} \vert {\dot {\tilde {\vec \psi}}} > &(50b). \cr}
$$

\noindent
A comparison of Eq. (49) and (40) demonstrates the EOM invariance.

Now, using Eqs. (41) and (44), it is easy to see that

$$
{\bf D}^{(\psi)} = {\bf C}^{\dagger} 
    \big[ {\bf D}^{(\chi)}\ {\bf C} + {\bf S}^{(\chi)}\ {\dot {\bf C}} \big]
\eqno(51)
$$

\noindent
Requiring that ${\bf D}^{(\psi)} = 0$ results in an equation which
defines ${\bf C}$

$$
{\dot {\bf C}} + 
{\bf \Gamma}^{(\chi)} \ {\bf C}  = 0 
\eqno(52)
$$

\noindent
where 

$$
{\bf \Gamma}^{(\chi)} \equiv {\bf S}^{{(\chi)}^{-1}} {\bf D}^{(\chi)} 
\eqno(53)
$$

\noindent
In principle, ${\bf \Gamma}^{(\chi)}$ is antihermitian since
${\bf S}^{(\chi)}$ is hermitian and ${\bf D}^{(\chi)}$ is
antihermitian.

\par\vfill\eject

\vskip 30pt

\centerline{\bf III.\ USE\ OF\ THE\ ISOP\ ALGORITHM}

\vskip 20pt

The ISOP algorithm[4] has been applied to several problems [4,5]
including the integration of two coupled equations resulting
from a time-dependent Hartree ansatz applied to $H_2$.[5]  
The ISOP algorithm is a two-time-step algorithm and involves (as
indicated above) a time interval $\Delta t$, the retarded
time (R), at the R-end of $\Delta t$, and the advanced time
at the A-end. The present equations are nonlinear 
equations--this presents no essential
difficulty however. It is possible, as we demonstrate below,
to decompose the solution into steps so as to in effect
render the equations linear from a computational point of view.
The basic computational tool used in taking time derivatives
in the ISOP is the Cayley formula.[7]  The superscripts
labeling $\pm$ and the orbital types are dropped in the
following.

First, rewrite Eq. (14) in matrix form as

$$
{\dot {\vec A}}  = -i {\bf H} 
    {\vec A}
\eqno(54)
$$

\noindent
Then, the analysis easily gives

$$
{\vec A}_A \approx
\big[ {\vec {\bf 1}} + {{i}\over{2}}\ dt\ {\bf H}_R \big]^{-1}
\big[ {\vec {\bf 1}} - {{i}\over{2}}\ dt\ {\bf H}_R \big]\ 
{\vec A}_{R}
\eqno(55)
$$

\noindent
where the $A$-subscript stands for advanced (in time) and the
$R$-subscript stands for retarded (in time), as has already
been used above.  Also,

$$
{H}_{k,j} = << \Phi_{k} \vert
   {\hat H} \vert \Phi_{j} >>
\eqno(56)
$$

\noindent
and then

$$
{\dot {\vec A}} \approx \big[ 
{\vec A}_{A} - {\vec A}_{R}  \big] / \Delta t .
\eqno(57)
$$

\noindent
The time derivative of the orbitals is evaluated in this manner also;
e.g.  for the $\chi$-set

$$
{\dot {\vec \chi}} \approx \big[ 
{\vec \chi}_{A} - {\vec \chi}_{R}  \big] / \Delta t .
\eqno(58)
$$

\noindent
The time advance of the matrix $\bf C$ may be evaluated (see Eq. (52))
in a similar manner (when needed) via

$$
{\bf C}_A \approx
\big[ {\vec {\bf 1}} + {{1}\over{2}}\ dt\ {\bf \Gamma}_R \big]^{-1}
\big[ {\vec {\bf 1}} - {{1}\over{2}}\ dt\ {\bf \Gamma}_R \big]\ 
{\bf C}_{R}
\eqno(59)
$$

It is useful to express the $H$-matrix elements (Eq. (55)) in terms of
the ${\hat h}$-matrix elements.  To do so, define

$$
h_{p(qr)s} \equiv
< \psi_p \vert {\hat h}_{qr}
   \vert\ \psi_s >
\eqno(60)
$$

\noindent
Then, (with $H_{jk} = H_{kj}$),

$$
\eqalign{
H_{11} &= h_{1(11)1} \ ; \cr
H_{12} &= {{1}\over{\sqrt{2}}} \big[ 
    h_{1(11)2} +  h_{1(12)1} \big] \ ; \cr
H_{13} &= h_{1(12)2} \ ; \cr
H_{22} &= {{1}\over{2}} \big[ 
    h_{2(11)2} +  2\ h_{2(12)1} +
           h_{1(22)1}
                                                     \big] \ ; \cr
H_{23} &= {{1}\over{\sqrt{2}}} \big[ 
    h_{1(22)2} +  h_{2(12)2} \big] \ ; \cr
H_{33} &= h_{2(22)2} \ ; \cr
H_{11} &= {{1}\over{2}} \big[ 
    h_{2(11)2} -  2\ h_{2(12)1} +
           h_{1(22)1}
                                                     \big] \ . \cr
}
\eqno(61)
$$

In order to use the ISOP to integrate the orbital equations, the
singlet and triplet cases are considered separately again.
The ISOP may now be applied directly to Eq. (19).  Using the
definitions of ${\hat T}_0(j)$ and $V_0(j)$ immediately below Eq. (3),
Eq. (19) may be written as (the explicit reference to particle
coordinates is dropped)

$$
{\dot {\vec \Psi}}  =
    - i\ {\hat T}_0  {\vec \Psi}
    - i\ {\hat {\bf M}}\ {\vec \Psi}
\eqno(62)
$$

\noindent
where

$$
{\hat {\bf M}} = {\vec {\bf 1}} V_0 + {\hat {\bf h}}_0
   + {\bf V}
\eqno(63)
$$

\noindent
such that ${\hat {\bf h}}_0$ is the same as ${\hat {\bf h}}$
in Eq. (23), except with ${\hat H}(1,2)$ replaced by ${\hat H}_0(1,2)$ (see
Eq. (3)).  Also ${\bf V}$ is the same as ${\hat {\bf h}}$ except
$V(1,2)$ replaces ${\hat H}(1,2)$.

Then, finally, application of the ISOP results in

$$
{\vec \Psi}_A = 
{{1 - {{1}\over{4}} i dt\ {\hat T}_0}
\over
{1 + {{1}\over{4}} i dt\ {\hat T}_0}}\ 
[{{\hat {\bf M}}_{bot}}]^{-1}\ 
 [{\hat {\bf M}}_{top}]\ 
{{1 - {{1}\over{4}} i dt\ {\hat T}_0}
\over
{1 + {{1}\over{4}} i dt\ {\hat T}_0}}\ 
{\vec \Psi}_R
\eqno(64)
$$

\noindent
where

$$
\eqalign{
{{\hat {\bf M}}_{bot}} &\equiv
+ {{1}\over{2}} i dt\ {\hat {\bf M}} \cr
{{\hat {\bf M}}_{top}} &\equiv
- {{1}\over{2}} i dt\ {\hat {\bf M}} \cr}
\eqno(65)
$$

Assuming that the $\chi$-set is orthonormal and that $\bf D^{(\chi)} = 0$,
as would be the case for a completely accurate time propagation
and spatial derivative and integral evaluation, the basic algorithm
proceeds as follows (reintroduce the orbital labels):

\vskip 10pt

\centerline{Basic\ Algorithm}

\vskip 10pt

\item\item{Step 1:} specify ${\vec \chi}_{R}(t=0), {\vec A}^{(\chi)}_{R}(t=0) 
\equiv 1$; 
\item\item{Step 2:} calculate ${{\bf a}^{(\chi)}_{R}}$ using Eqs. (22,23);
\item\item{Step 3:} calculate ${{\hat {\bf h}}}^{(\chi)}_{R}$ using Eq. (37c);
\item\item{Step 4:} calculate ${{\bf H}}^{(\chi)}_{R}$ using Eqs. (59);
\item\item{Step 5:} calculate ${\vec A}^{(\chi)}_{A}$ using Eq. (55);
\item\item{Step 6:} calculate ${{\vec \Psi}^{(\chi)}}_{R}$ using Eq. (39b);
\item\item{Step 7:} calculate ${{\vec \Psi}^{(\chi)}}_{A}$ using Eq. (62);
\item\item{Step 8:} calculate ${{\bf a}^{(\chi)}_{A}}$ using Eqs. (22,23);
\item\item{Step 9:} calculate ${\vec \chi}_{A}$ using Eqs. (37b) after
having taken the inverse of ${{\bf a}_{A}^{(\psi)}}$;
\item\item{Step 10:} calculate ${\dot {\vec \chi}}$ using Eqs. (58);
\item\item{Step 11:} calculate ${{\bf S}}^{(\chi)}_{R}$ using Eq. (25);
\item\item{Step 12:} calculate
${\bf D}^{(\chi)} \equiv
    < {\tilde {\vec \chi}} \vert\ {\tilde {\dot {\vec \chi}}} >$; 
\item\item{Step 13:} if ${{\bf S}}^{(\chi)}_{R} \approx {\bf 1}$
and if ${{\bf D}}^{(\chi)} \approx {\bf 0}$, continue, otherwise, 
jump out of this algorithm;
\item\item{Step 14:} let the advanced time become the retarded time for
another time interval and then go to step 3 and continue.

\vskip 10pt

If an exit of the basic algorithm occurs because of the lack of
orbital orthonormality or the lack of a null $D$-matrix, the
\lq\lq Correction Algorithm\rq\rq\ is implemented, for
the particular $\Delta t$ in question, as follows:

\vskip 10pt

\centerline{Correction\ Algorithm}

\vskip 10pt

\item\item{Step 1:} assume ${{\bf C}}_{R} = {\bf 1}$;
\item\item{Step 2:} calculate ${{\bf \Gamma}}_{R}$ using Eq. (53);
\item\item{Step 3:} calculate ${{\bf C}}_{A}$ using Eq. (59);
\item\item{Step 4:} calculate ${\vec \psi}_{A}$ using Eqs. (41);
\item\item{Step 5:} calculate ${{\bf a}_{A}^{(\psi)}}$ using Eq. (46);
\item\item{Step 6:} calculate ${{\hat {\bf h}}}^{(\psi)}_{A}$ using Eq. (48);
\item\item{Step 7:} let the advanced time become the retarded time for
another time interval and then go to step 4 of the 
\lq\lq Basic Algorithm\rq\rq\ and continue (letting $\psi \rightarrow \chi$).

\noindent
Thus, the solution may be efficiently propagated in time in a way
that guarantees orbital orthonormality and a null $D$-matrix
(non-reduntant time evolution).

\vskip 30pt

\centerline{\bf IV.\ CONCLUSIONS}

\vskip 20pt

Now returning to the TDHF[9], the one-electron 
orbitals are assumed orthonormal
and the derivative matrix elements are 
set to zero.  This is possible in the
context of the TDHF because the EOM are derived
using the Dirac-Frenkel variational principal (DFVP).[10,11]  In fact,
orthonormality of the one-electron orbitals and zero $D$-matrices are
not restrictive in the context of the TDHF.  
This can be demonstrated
by a set of guage transformations in the 
manner of Refs. (5) and (9). The key idea of the
TDHF derivation using the DFVP is the 
independent variation of the one-electron
orbitals.[9]  The authors of Refs. (2) and (3) 
also use the DFVP to 
derive their EOM for a Hartree-like ansatz 
which is used for distinguishable
particles.  The present derivation of a set of exchange equations proceeds
in the manner of Ref. (5)--that is by projections.  It is necessary to
have a way of determining the $D$-matrices. This is so
because the time derivative
that appears inside the spatial integral is at the same time-step as the
time-derivative of the Schr\"odinger equation itself.  This results
in the necessity of a self-consistent procedure at each time-step--an
operation that needs to be avoided if possible.  If projections are
used to derive the two coupled equations for the one-electron orbitals,
the $D$-matrices can not be set to zero 
without a \lq\lq hidden symmetry\rq\rq\ analysis similar to that given
above.  This in fact has been done.[12] The equations that result from
that procedure are
distinctly different from the TDHF equations.  

The present work is an
alternative derivation of a set of time-dependent exchange equations (TDEEs),
which follows immediately from the MCTDH theory of [2,3].  The present
equations are derived by a projection procedure, but are exactly the
same equations that would be produced by the DFVP.  This all results
from the reduntancy which follows from the use of the purely time-dependent
coefficients multiplying each two-electron configuration.

An algorithm for the solution of the new exchange equations has been
given using the ISOP method and a procedure was described to enforce
orbital orthonormality and a null $D$-matrix.  This allows for
a non-reduntant time evolution.

\vskip 40pt

\centerline{\bf ACKNOWLEDGEMENTS}

\vskip 20pt

The research was supported by the Army High Performance
Computing Research Center and the US Army, Army Research
Laboratory (DAAH04-95-2-0003/ contract number DAAH04-95-C-0008),
by NSF CREST grant HRD-9707076, and by the Lawrence Livermore
National Laboratory Research Collaboration Program for
Historically Black Colleges and Universities and Minority
Institutions.  
The author would like to acknowledge
useful conversations with  H.-D. Meyer, Burke
Ritchie, Merle Riley and Mario Encinosa.

\par\vfill\eject

\vskip 30pt

\centerline{\bf REFERENCES}

\vskip 20pt

\item{ [1] }C.F. Barnett, \lq\lq Atomic 
Collision Properties\rq\rq\ in
$\underline{A\ Physicist's\ Desk\ Reference:}$ \hfill\break
$\underline{The\ Second\
Edition\ of\ Physics\ Vade\ Mecum}$, American 
Institute of Physics, New
York, N.Y., ed. H.L. Anderson (1989), p. 92.
\item{ [2] }H.-D. Meyer, U. Manthe, and L.S. Cederbaum,
Chem. Phys. Lett. ${\bf 165}$, 73 (1990).
\item{ [3] }U. Manthe, H.-D. Meyer, and L.S. Cederbaum,
J. Chem. Phys. ${\bf 97}$, 3199 (1992).
\item{ [4] }B. Ritchie and M.E. Riley, Sandia Report Sand97-1205,
UC-401 (1997).
\item{ [5] }B. Ritchie, C.A. Weatherford, International J. Quant.
Chem. S${\bf 70}$, 627 (1998).
\item{ [6] }E.K.U. Gross, E. Runge, and O. Heinonen,
$\underline{Many-Particle\ Theory}$, Adam Hilger, New York (1991).
\item{ [7] }M.D. Feit, J.A. Fleck, and A. Steiger,
J. Comput. Phys. ${\bf 47}$, 412 (1982).
\item{ [8] }A. Szabo and N.S. Ostlund,
$\underline{Modern\ Quantum\ Chemistry}$, McGraw-Hill, 
New York (1989), pgs. 142-145.
\item{ [9] }A.K. Kerman and S.E. Koonin, Ann. Phys. (N.Y.)
${\bf 100}$, 332 (1976).
\item{ [10] }P.A.M. Dirac, Proc. Cambridge Philos. Soc. ${\bf 26}$,
376 (1930).
\item{ [11] }J. Frenkel, $\underline{Wave\ Mechanics:\ Advanced\ General\
Theory}$, Clarendon (Oxford) (1934).
\item{ [12] }M. Riley, B. Ritchie, and C.A. Weatherford, unpublished work.

\par\vfill\eject

\end